\documentclass[11pt,letterpaper, twocolumn]{article}

\usepackage[utf8x]{inputenc} 
\usepackage{fullpage}
\usepackage{setspace}
\usepackage[usenames]{color}
\usepackage{url} 
\usepackage{mdwlist} 
\usepackage{hyperref}
\usepackage[sf,bf]{caption}
\usepackage[bf,compact]{titlesec}
\usepackage{graphicx}
\usepackage{multirow}

\usepackage[T1]{fontenc}
\usepackage{times}
\usepackage[scaled=0.9]{helvet}

\begin{document}

\title{The BitTorrent Anonymity Marketplace}
\author{
Seth James Nielson,  Dan S. Wallach \\
{\small seth@sethnielson.com, dwallach@cs.rice.edu} }

\date{Computer Science Department, Rice University}
\maketitle

\begin{abstract}
The very nature of operations in peer-to-peer systems such as BitTorrent
exposes information about participants to their peers. 
Nodes desiring anonymity, therefore, often chose
to route their peer-to-peer traffic through anonymity relays, such as Tor.
Unfortunately, these relays have little incentive for contribution and struggle
to scale with the high loads that P2P traffic foists upon them. We propose a novel
modification for BitTorrent that we call the \textit{BitTorrent Anonymity Marketplace}.
Peers in our system trade in $k$ swarms obscuring the actual
intent of the participants. But because peers can cross-trade torrents, the $k-1$ 
cover traffic can actually serve a useful purpose. This creates a system wherein a
neighbor cannot determine if a node actually wants a given torrent,
or if it is only using it as leverage to get the one it really wants.
In this paper, we present our design, explore its operation in simulation, and analyze its effectiveness.
We demonstrate that the upload and download characteristics of
cover traffic and desired torrents are statistically difficult to distinguish.
\end{abstract}

\section{Introduction}
\label{intro}
Peer-to-peer file transfer protocols, such as the very popular BitTorrent~\cite{bittorrent-incentives} protocol, 
provide massively scalable architectures for distributing large files. 
Unfortunately, privacy is a direct casualty of the peer cooperation that drives them. For traditional
client-server architectures, the client need only trust the server not to reveal 
to additional parties the details of the transaction. While some information is revealed
just from observing that the client and server communicated with each other, the 
specifics are confidential. With appropriate cryptographic and protocol mechanisms,
the client can have strong assurances of privacy in the transaction so long as the server
remains trusted.

On the other hand, in peer-to-peer cooperation, an individual, by necessity, reveals
details of the transaction to many parties, each of which must be trusted if privacy is
to be maintained. This problem is exacerbated by the nature of peers in such systems.
In the client-server model a user can limit interactions to well-known, vetted
servers, but in contemporary p2p systems peers could be controlled by an incompetent
or malicious individual or organization.

A number of solutions to the peer-to-peer anonymity problem have been proposed.
The most common solution in practice is to route traffic through anonymity relays such as
Tor~\cite{Dingledine04Tor}. Unfortunately, Tor has, by default, no incentives for cooperation
and struggles to scale with P2P workloads. Our goal at the onset of this research was to
develop an anonymity mechanism for BitTorrent that incentivizes participation
and induces scalability. Such a mechanism would provide better performance than
BitTorrent-over-Tor while still providing sufficient anonymity guarantees. Furthermore, it
would draw BitTorrent users away from the Tor network and all parties would be better off.

We have created the \textit{BitTorrent Anonymity Marketplace} as novel solution to this problem.
This system provides genuine incentives for nodes to participate
in cross trading of multiple swarms obscuring the actual intent of the driving nodes creating what we refer to as
\textit{k-traffic-anonymity}. We demonstrate in simulation the effectiveness of
this obfuscation and show that it has nearly optimal performance tradeoffs. Our result is distinguished from other BitTorrent
specific anonymity solutions either because participation is incentivized, or because the attack model
we address is more powerful. 

This paper is organized as follows. We first review some of the operations of BitTorrent and some of the
principles of incentives in Section~\ref{background}. In Section~\ref{related} we review the current solution
space to the p2p anonymity problem. Then we introduce our own objectives and design in Section~\ref{objectives}. 
We evaluate our results in Section~\ref{evaluation}. 
Finally, we close with a discussion of our research in Section~\ref{discussion} and our conclusions in
Section~\ref{conclusion}.

\section{Background}
\label{background}
\subsection{BitTorrent}
BitTorrent~\cite{bittorrent-incentives} is a highly successful and 
popular peer-to-peer protocol that enables efficient, 
rapid distribution of potentially large amounts of data to a group 
of clients. It utilizes the available upload bandwidth 
of the participants to scale to support many 
users. Most important, it has built-in incentives mechanisms that
reward correct participation.

To make an item available for BitTorrent downloading, a publisher 
makes available a \emph{tracker} and at least one \emph{seed}. The tracker
follows the nodes participating in the swarm, helping nodes locate their peers.
Seed provide round-robbin, best-effort service to all connecting peers.

To download the object, a group of nodes, collectively called the \emph{swarm} 
join the system by contacting the tracker, indicating their intent to participate.
The tracker informs joining nodes of random subsets of their peers. The nodes
then establish direct connections with these subsets forming their local 
\emph{neighborhoods}. Thus joined, the nodes download the object in equal sized chunks of the file called
\emph{pieces}. Nodes share information with their neighborhood about the pieces
they have available and update them as new pieces are acquired.
 
Nodes, however, limit the number of peers in their neighborhood that can download
from them at any given time. They evaluate their peers based on how much each has recently uploaded.
The node then provides download service to the top three or four contributors. 
Each node also provides 
 service to one or two random nodes as a method of searching the neighborhood for better partners. Thus, peers
 have an incentive to contribute to their neighbors in order to receive reciprocal contributions from their neighbors
 in turn. When a node decides to service a peer, it is said to \emph{unchoke} the peer. Conversely, when it will
 no longer serve the peer, it is said to \emph{choke} it. Once a peer is unchoked, it can send
 \emph{Request} messages asking for data. If the unchoking node refuses, the peer
 considers itself snubbed and will not do business with that node for some time. Nodes update
 their peers with \emph{Have} messages when a new piece is received so that the neighborhood
 keeps abreast of what a node can and cannot trade.
 
 While a significant corpus of research has demonstrated that BitTorrent can be 
 exploited~\cite{bittyrant1_2007,tian,sirivianos,liogkas}, BitTorrent continues to
 work well in practice. The incentives in BitTorrent are sufficient, at present, for 
 keeping the system stable. Indeed,  while there is no consensus on the true
amount of BitTorrent data in-flight today, it is clear that the number is
large at somewhere between one-third and one-half of all Internet 
traffic~\cite{bittorrent_traffic_original, torrentfreak_myth, wired_bittorrent_traffic,
ipoque_bittorrent_traffic}.

\subsection{Incentives}
Peer-to-peer systems' greatest strength is their lack of centralization. At the same time, this lack of
centralization makes enforcement of peer behavior difficult. In general, the system designers
intend for peers to behave in a certain way, but peers may choose to behave differently.
Most nodes are assumed to be \textit{rational}, or self-interested, and want to maximize their benefit from the system while
simultaneously minimizing their own contributions. \textit{Faithfulness}
is the measure of a node's obedience to designer specification. By definition, rational nodes
are only faithful if it is in their own interest, and, therefore, faithfulness can only be achieved
by designing systems with proper incentives~\cite{faithfulness-formal, faithfulness-applied}.

In previous work, we identified two general classes of incentives in peer-to-peer systems:
artificial and genuine~\cite{Nielson05Taxonomy}. Genuine incentives are characterized by
being an intrinsic property of the p2p protocol, whereas artificial incentives are a super-imposition
of reward and punishment on top of an unincnentivized system. The intrinsic nature of genuine
incentives makes them more robust to rational manipulations and are, therefore, preferred.

\section{Related Work}
\label{related}
A number of solutions to the peer-to-peer anonymity problem exist or have been
proposed. We briefly outline some of these approaches here.


\subsection{Tor}
Tor~\cite{Dingledine04Tor} is distributed network of relays operated by volunteers that allows clients to route
network traffic through them to disguise the true origin. If used properly, the client's identity
and physical location are kept hidden from other entities.
Per-relay encryption also provides anonymity against wire-traces and packet sniffing.
Each relay is allowed to define its own
policy about what it will and will not do for the network. \textit{Entry routers}, as the name implies, 
accept traffic from outside the Tor network. Conversely,
\textit{exit routers} allow traffic out to the true destination. \textit{Middle routers} only relay traffic within Tor itself. 

A node that desires anonymity computes an \textit{onion route} through the Tor network.
It encrypts its packet with a layer of encryption for each router in the network.
Each intermediate node peels off a layer of encryption, and forwards the traffic to the next hop.
Each node only knows the preceding and subsequent steps in the route. The nodes cannot be sure
if the packet they are receiving is from the original sender, or simply a relay in the route.

Measurements taken in~\cite{understanding_tor_2008} indicate that 40\% of the traffic from a sample Tor exit node
was used for BitTorrent indicating how popular Tor is for providing BitTorrent anonymity.

Despite Tor's usefulness, it does struggle with a significant problem. It has trouble 
encouraging participants to contribute new computers to serve in the Tor network,
impacting Tor's ability to scale with the traffic it receives.  Additional nodes also
strengthen anonymity.  
However, the value of serving as a relay to a user is unclear; it has no impact on the
quality of service that they observe from the Tor network.  
Consequently, most users choose not to contribute.

Another important observation is that any negative legal or social response resulting from the originator's
connection will be borne by the exit node. Consequently, many nodes have a strict disincentive to 
not serve as an exit node.

\paragraph{Artificial Incentives for Tor} Recently, researchers have proposed extending Tor 
with incentives for better participation. One proposal~\cite{incentives_in_tor2008} is to
create a central authority that tracks each node's contributions and publicizes their good behavior so
that other nodes can reward them. Alternatively, other research proposes micropayments,
where Tor users may buy a higher quality of service~\cite{columbia_incentives_in_tor2008}.

\subsection{BitTorent Specific Solutions}
In addition to the Tor general anonymity network, anonymity mechanisms have been proposed
that are specific to BitTorrent.

\paragraph{BitBlender} \cite{bitblender2008} extends
BitTorrent to route traffic through peers in an anonymity directory. In a fashion similar
to Tor, members of the swarm can forward requests through other peers providing
a form of anonymity it calls ``$k$-anonymity.''  They define this as
``users are indistinguishable from a set of $k$ users.''  Unfortunately, as with Tor, BitBlender
provides no incentive for nodes to offer relay services. Please note that
$k$-anonymity in their system is not the same as $k$-traffic anonymity in our paper.

\paragraph{OneSwarm} \cite{oneswarm2009} attempts to solve the BitTorrent anonymity problem more
generally. Nodes have extensive control
over what information about themselves they will share and with whom. In particular, OneSwarm would
be used with social networking so that information is only shared with ``friends.'' OneSwarm does not
solve the problem of maintaining anonymity in large groups of untrusted peers.

\paragraph{SwarmScreen} \cite{swarmscreen2009}, in a fashion similar to our work,
proposes anonymity through
the use of cover traffic. In particular, they assert that nodes achieve plausible deniability
``by simply adding a small percent (between 25 and 50\%) of additional random 
connections that are statistically indistinguishable from natural ones.'' SwarmScreen's attack model
has an observer classify nodes based on the behavior of the community they participate
in. Its stated goal is the disrupting of these ``guilt-by-association'' attacks,
or in other words, obscuring the \textit{community} that a node is participating with at any given point in time.
We will make further comparisons to SwarmScreen as we outline our own solution. Our work is
only superficially similar.

\section{Design}
\label{objectives}
Our objectives for this work break down into three categories: anonymity, performance, and incentives. 
As we detail our objectives, we will compare and contrast our solution with SwarmScreen to illustrate
the differences in approach and philosophy.

Our primary goal is an obfuscation of participant behavior that we call 
\textit{k-traffic-anonymity}. Nodes in our system must have an indistinguishability of
intent as they are observed by their peers. In other words, a node's peers can see that they
are downloading $k$ items but cannot distinguish
which one of them the node picked intentionally. The intentionally picked torrent is called
the \textit{native interest}.

Our primary threat: observers wish to ascertain a target node's native interest.
We call the attacker an \textit{inquisitor} and define three different classes of attacks. 
\textit{Fully passive}
inquisitors do not contact any other peers. Instead, these nodes
exclusively scan the tracker's data on where nodes are participating.
\textit{Decoy passive} inquisitors do contact peers
and can appearance to participate. 
They may lie and announce piece reception, make requests for pieces from
their peers, and in any other way appear to be normal nodes, but they will not actually 
accept downloads or make uploads. Finally, \textit{Active} inquisitors can participate and
behave like any other node in the system. 

Within our anonymity constraints, we want good performance. We will measure
performance in terms of the number of additional download bytes 
required to achieve a given level of anonymity. In an idealized
world 
where all torrents are the same size, optimal performance for $k$-traffic-anonymity is
$k$ times the number of bytes in a torrent. In other words, the node downloads exactly $k$ torrents
and nothing more. Our objective is nearly optimal performance; we are not interested in designs,
for example, that require $2k$ or more download cost for $k$-traffic-anonymity.

Finally, our last objective is that the incentives structure of our system encourages full participation of
the rational nodes. The critical incentive that we identify is that \textit{
participating in a torrent, purely for anonymity reasons, can still offer performance benefits}.
This is important for two reasons connected with anonymity.
First, to do otherwise would create a system wherein some torrents might only ever have natively interested
nodes downloading it. This is a form of anonymity starvation. 
Second, if there is no value to the cover-traffic torrents in the download set, 
an inquisitor might be able to distinguish the native-interest in the set.
By creating a system where all torrents can be valuable
as cover-traffic, nodes have incentives to participate in them preventing torrent starvation and obscuring the
native interests of the participants. We emphasize that this is a genuine incentive, requiring no additional
enforcement mechanisms or auditing.

In contrast, SwarmScreen is interested in a much weaker attack model. They 
showed that BitTorrent communities tend to form around \textit{interests} rather than around language, geography,
or even friendship. They further showed that these communities can be monitored and classified by observing
a small number of the nodes. The describe this invasion of privacy as ``guilt by association'' attacks. Finally,
the also demonstrated  that monitoring
just 1\% of the network is sufficient for assigning users to their communities with 86\% accuracy. They solve
this attack model by mixing in traffic to other random torrents to obscure which community a SwarmScreen participant
belongs to. Defeating this simpler attack model only costs them 25\% to 50\% overhead.

However, the stronger attack model we defeat with our system is worth the increased cost. An observer
that can follow a SwarmScreen node for a long period of time can easily determine which torrents
the node was downloading, because the node never fully downloads the torrents it uses as cover traffic.
At the same time, our system also disrupts the guilt by association attack as described.

\paragraph{BitTorrent Anonymity Marketplace, High-Level Design.}
Our basic system works for any given $k$ level of anonymity.
First, each node participates in $k$ different torrents simultaneously. It advertises all $k$
torrents, hereafter called its \textit{active set}, to its local neighborhood. While the composition
of the active set can change over time, it must eventually completely download $k$ complete 
torrents (we will call these the \emph{download set}), or else a long-term observer could immediately filter out the cover-traffic.

Our design also requires that nodes will ``cross trade'' their torrents, i.e., a node unchoke its peers' requests for {\em any}
torrent, not just the torrents where a node has benefited from its peers.  In our design, a node will consider every possible
torrent it sees advertised by its peers, and will prefer to join those torrents which it believes will be most beneficial
in its quest to download its native interest.

The design of our valuation function is drawn from models of supply and demand in
economics~\cite{wealth_of_nations}. In general, the value of a torrent to a node is raised by increased numbers
of peers that desire it, while the value is lowered by increased numbers of peers that provide it. Unfortunately,
it is impossible to directly measure a torrent's supply and demand in BitTorent, and so
we use several factors to approximate this. These factors include how much of the torrent the peer requires to complete it,
\emph{Have} announcements indicating what it is currently trading, and direct \emph{Request} messages to measure
what is available.

We highlight that our valuation function was derived from empirical data and not an economic
or mathematical model. Developing a coherent economic valuation function is a significant research undertaking
in and of itself and is beyond the scope of this paper. Our experimental version was constructed
by taking the factors that impact the value of a torrent and combining them in a weighted sum. This
construction, similar to how utility functions are built~\cite{utility_frombook}, enabled us
to experiment with different weights for the factors by dialing up or down the constant associated with that variable.
Later in this paper, we will detail our derivation of our constants from experimentation.

The critical hypothesis tested in this work is whether using a valuation function on torrents will drive node
behavior such that protocol exchanges related to the native interest are statistically indistinguishable from
protocol exchanges for cover traffic. The core idea is that a peer has no idea if a node is asking for pieces
of a torrent because it actually wants it, or if it is just asking for those pieces because it has 
a high value due to the neighborhood's ``market'' conditions.


\section{Evaluation}
\label{evaluation}

We employed a simulator developed in previous research~\cite{longterm}
to evaluate our implementation of the BitTorrent Anonymity
Marketplace. The simulator, running faster than real-time, enabled
fast design cycles. After completing a simulation, we studied the
results, modified the configuration, and re-ran our experiments. This
was a significant advantage over using an artificial environment such
as PlanetLab or EmuLab 
to run a ``real'' BitTorrent client. Simulation is also preferred to
releasing a client to public users because it allows us better access
to system and client state information and it avoids any potential
legal or ethical issues we are not yet prepared to confront.

\subsection{Implementation}

Our client implementation was developed to be as realistic as possible
in all stages of their operation. One notable departure from a stock
BitTorrent system is that we assume the presence of a distributed hash
table (DHT) in which to store metadata, rather than the more limited
tracker functionality in the current BitTorrent. What follows is an
overview of how nodes participate in the Marketplace.

\paragraph{Publishing.} It is essential that objects exchanged in the Marketplace are 
opaque to users that are uninterested in them. Otherwise, users may
choose not to trade in objects they deem overly sensitive. For this
reason, all content is encrypted and assigned random identifiers.
We assume out-of-band methods (e.g., publisher web servers)
help users discover specific torrents and obtain the decryption keys.
In this manner, participating nodes will
trade in many torrents without any knowledge of their content, except
for their own native interest, thus obtaining a modicum of plausible deniability. 
Once a publisher has encrypted the object and created its random
ID, it stores a record similar to a torrent-file into the
DHT and announces nodes that are seeding the torrent within the DHT.

\paragraph{Messages.} All inter-peer communication consists of unmodified 
BitTorrent messages with one exception. While normal BitTorrent \emph{Choke} and
\emph{Unchoke} messages identify a specific torrent, 
in the Marketplace these messages are 
not torrent-specific. These two messages
instead signal that the sender is willing or unwilling to fulfill requests
for any of the torrents it has currently advertised.

\paragraph{Joining.} To use the BitTorrent Anonymity Marketplace, a participant first
acquires the random ID for the desired object, as described earlier.
Next, the node joins the
DHT and requests a list of active torrents. From this list, the node creates
a list of $k$ torrents consisting of its desired torrent plus $k-1$ randomly selected torrents.
The node then indicates to the DHT that it is joining those
$k$ torrents and requests participating peers. The node creates a neighborhood
from these lists, preferring peers that show up in multiple torrents.

\paragraph{Trading.}  After nodes join the system, they unchoke peers in a manner similar to BitTorrent
with the highest upload services getting the unchoke slots. However, in the Marketplace,
all upload service is adjusted by the estimated value of the received pieces. 
Our implementation keeps the value constant across an entire torrent,
although different pieces could ostensibly have different values. Once the values
of the upload services are adjusted, unchoking proceeds normally. At the same time,
if the node can find a more valuable torrent than the least valuable torrent in its active 
set, it drops that torrent and joins the new one.

\paragraph{Seeding and Termination.} A Marketplace participant must complete $k$ downloads before leaving the system. 
Before all $k$ torrents have completed, a node may find value in seeding one of its completed torrents,
depending on its observations of the supply and demand for those torrents.  Alternately, it could forgo seeding
and instead look for more profitable ways to trade its available bandwidth.


\subsection{Development of the Valuation Function}

We have developed a valuation function based on reasonable economic assumptions,
refined by experimentation, and suitable for enabling our evaluation of our anonymity objectives.
We started with basic supply and demand concepts~\cite{wealth_of_nations}. In other words, we accept the 
assumption that increased desire and scarcity raise the value of a given object,
while decreased desire and abundance reduce the value of same. In terms of the BitTorrent Anonymity 
Marketplace, the number of nodes wishing to download a pieces of a torrent constitute the desire,
and the nodes that can service those requests constitute the supply. These two factors 
are the basis for our valuation function.

Unfortunately, the node cannot measure these factors directly and must therefore estimate them.
For example, a node sees all the peers within its neighborhood, but it cannot see further.  It cannot
see every peer participating in every torrent, thus it cannot determine the global supply and demand
of torrent pieces, nor even can it determine any other peer's view of this data.
To estimate supply, Marketplace nodes
treat what they can see, within their own neighborhood, as an estimate for what their peers can see.
(Neighborhood visibility is not transitive.  If $A$ is in $B$'s neighborhood and $B$ is
in $C$'s neighborhood, there is no guarantee that $A$ knows anything about $C$.)
Nodes can make a better estimate about the demand for a torrent by totaling the number of pieces required
for their peers. They then combine these two estimates into a single
factor hereafter referred to as \textit{approximate demand over supply}.

In addition to this information, BitTorrent nodes can make use of the \emph{Have} announcements and \emph{Request}
messages from peers to know more about demand in the neighborhood. The \emph{Have} messages indicate a
degree of freshness to what torrents neighbors are trading and, of course, \emph{Request} messages are the
strongest, most straight-forward measure of demand available.

Our early valuation function was a weighted sum of these three factors. Using this construction, each factor 
could be experimentally measured to determine if it had an impact at all, and the ideal weighting could be
derived experimentally using
our simulations. By fixing a weight of $1$ to all but one factors, the remaining factor can be evaluated
independently. Setting
this experimental factor to $0$, for example, completely eliminates its impact on the function.

For testing the Marketplace, we fixed $k=5$, set the total number of torrents in the marketplace to 40,
initiated 100 clients plus 40 seeds, and used 125 MB files for each 
torrent\footnote{Individually 125 MB is a small file for BitTorrent, but because our nodes are exchanging five files
simultaneously, the amount of data in transit is 625MB per client.}. For simplicity, all the clients 
have the same upload and download speeds of 1Mbps, start at the same
time, and end when their $k$ downloads are complete. To test the effects of torrent popularity, we configured
10\% of the torrents to be the native interest of 50\% of the clients.

Our initial simulations immediately demonstrated that our initial valuation function was insufficient.
Regardless of configuration, the
clients in the simulation would not complete their downloads. We determined that the nodes
were dropping the torrents in their active set, regardless of how much they had completed, for a new torrent that
was surging in popularity in their neighborhood. We decreased the frequency at which nodes would
update their active set but that didn't solve the problem sufficiently.

After some additional experimentation, we determined that because one of the goals of the node is to complete $k$
downloads, the completeness of a torrent should factor into the valuation function. In other words, if all other
factors are equal, a more complete torrent should be valued higher than a less complete one. We retooled the
valuation function with this new factor and re-ran the simulations and were rewarded with converging results.

Using our more mature valuation function, we tested the factors in the function independently.
For each factor tested, we experimented with weights of 0, 0.25, 0.5, 0.75, 1.0, 2.0, 4.0, 8.0
and 16.0. For completeness, we also tested a few other non-value related variables such as how often the node
updates its active set, and so forth. In total, we ran fifty different different configurations of the simulator,
again fixing all but one factor at a time and varying it across this broad range of weights.

These tests demonstrated, again, that biases toward completing torrents that have been started are
essential, and that data collected from direct requests is the best proxy for overall demand.
When we reconfigured the simulation to ignore direct requests,
performance worsened by nearly twenty percent. Interestingly, the remaining factors proved to be
much poorer estimates of demand and had little impact on average performance. However, they
are useful to a node at times when the node has not recently received any such requests. A small
weight for these factors was better than no weight at all.
We conclude that when the direct request factor is in play, it should dominate the equation. However, when the
direct request factor drops to zero, these weaker factors serve as a backup.

While the specific coefficients of valuation function are optimized for our simulation configuration and are thus not directly
applicable for a real-world deployment, the
insights obtained from this empirical evaluation are still essential. Moreover, we can now
test our central hypothesis: will cross-trading nodes that 
use a valuation function to decide which cover-traffic nodes to trade
have the $k$-traffic-indistinguishability property? 

\subsection{Anonymity Results}

To evaluate anonymity, we took the best observed weight for each of the valuation factors and
reconfigured the simulator appropriately. With this valuation configuration, we ran twenty simulations.
Each took several hours to complete on a 2.4 Ghz Athlon and covered
approximately 7 hours of simulation time.
Each run involved about 70GB of simulated data transfer and approximately 10 million control messages.
The simulations output logs that detail the data transfers and control messages and we used
them to trace how the peers interacted with each other as well as to calculate costs and determine performance. 

Our primary goal was to quantify indistinguishability of intent. This property means that a node
downloading $1$ native interest, and $k-1$ cover traffic torrents will not reveal its native interest by
its behavior to its peers. We will examine three node behaviors that could potentially reveal
the native interest to peers: start times for torrents, end times for torrents, and download patterns.

\paragraph{Start Time.} We first evaluate the indistinguishability of start times, where start time is
measured as an integer rank. In other words, the first torrent that a node makes requests for is ranked $1$, 
and the second torrent that a node makes requests for is ranked $2$, and so on. We evaluated this
aspect of indistinguishability in two ways.

\begin{figure}
\centering
\includegraphics[scale=.25]{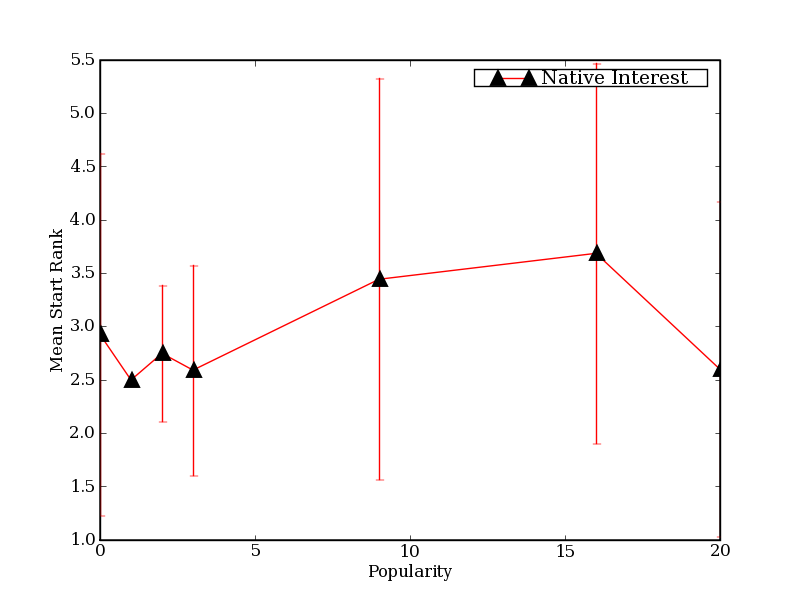}
\caption{The mean start rank of native interests plotted against popularity. 
The $x$-axis is the number of peers natively interested in the torrent, the 
$y$-axis is the starting rank. The error bars show the standard deviation. The 
wide standard deviations mean that native interests have a wide
range of start rank.\label{native_startrank}}
\end{figure}

First, we checked that there was sufficient variability of start times for native interests. It is important, of course,
that native interests not have a predictable start rank. Our results are shown in Figure~\ref{native_startrank}. 
The graph is parameterized on the number of nodes natively interested in the torrent, as a measure of popularity.
The $y$ axis is the average start rank for nodes of that popularity and the standard deviation. The graph shows
that the standard deviation is high for start rank, so a node's native interests are suitably obscured from its peers.

\begin{figure}
\centering
\includegraphics[scale=.25]{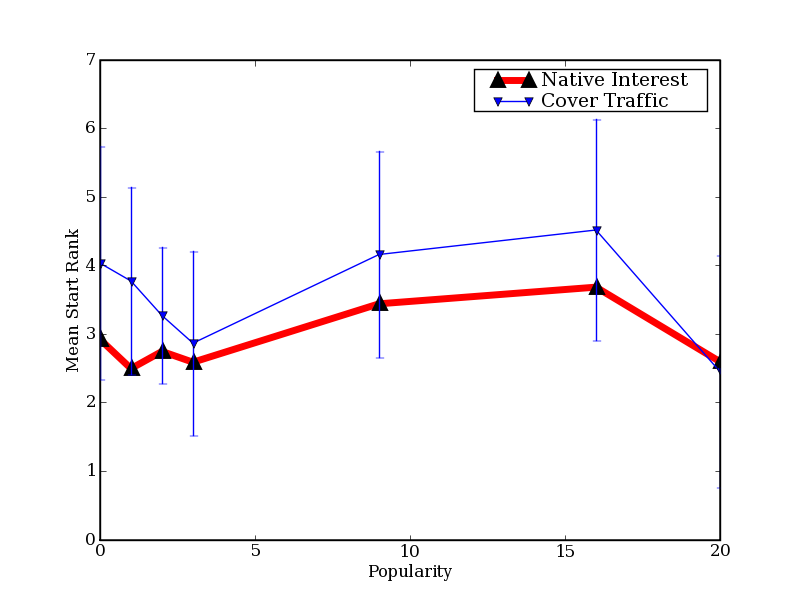}
\caption{The mean start rank of the various torrents plotted against 
the start rank for the same torrent for peers not natively interested. This graph shows
that native interests do start sooner, but the mean lies within the standard deviation 
of non-native interest start times for most torrents.\label{startranks_cmp}}
\end{figure}

Our second measure of the indistinguishability of start times is to measure the
average start time for the same torrent for peers that are natively interested relative to
peers that are not (see Figure~\ref{startranks_cmp}). 
There is a noticeable shift to earlier start times for native interests. Nevertheless, 
the average times for the native interests lie within the standard deviations of the
start times for non-native interests. The distributions are not statistically different enough
to be detectable. Furthermore, the native and non-native
graphs have similar shapes, suggesting similar behavior for the two populations.

\begin{figure}
\centering
\includegraphics[scale=.25]{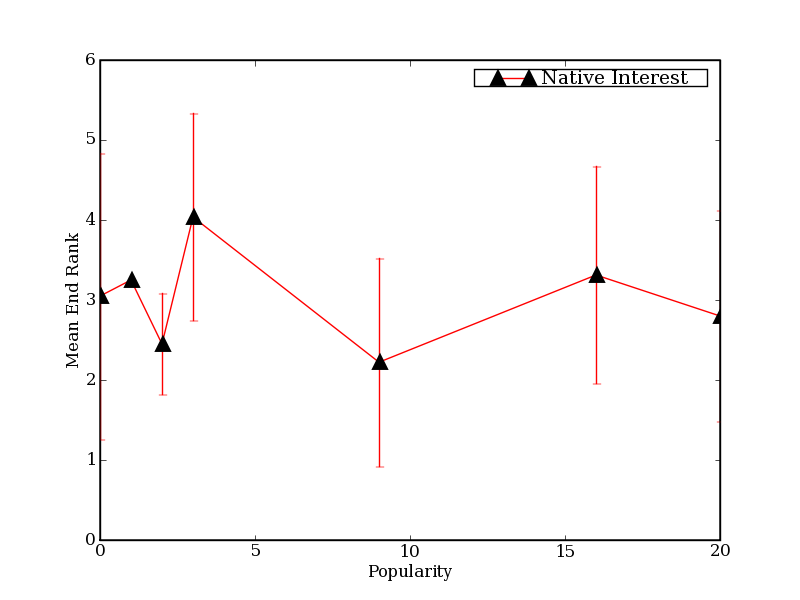}
\caption{Similar to Figure~\ref{native_startrank}, this graph shows
the mean ending ranks and the standard deviation. As with start times,
end times vary sufficiently to make them poor predictors of interest.
\label{native_endrank}}
\end{figure}

\begin{figure}
\centering
\includegraphics[scale=.25]{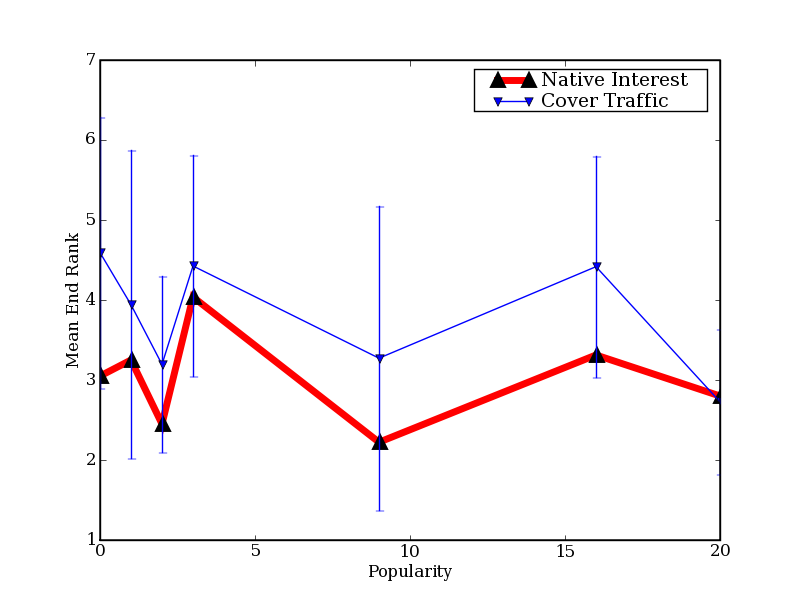}
\caption{The mean end ranks for native interest compare to mean
end ranks for non-native interest. As before, there is a noticeable shift
downwards, but, as before, the means for the native interests tend to 
lie within the standard deviations of the non-native interests.
\label{endranks_cmp}}
\end{figure}

\paragraph{End Time.} It is also important that native interests not end predictably.
Expressing end times as integer ranks, we evaluated the variability of native end times in Figure~\ref{native_endrank}
and compared those times to non-native end times in Figure~\ref{endranks_cmp}.
These graphs show that, as with start times, there is a wide variability in the
end times and that the mean is within the standard deviation of cover-traffic start times.

\begin{figure}
\centering
\includegraphics[scale=.30]{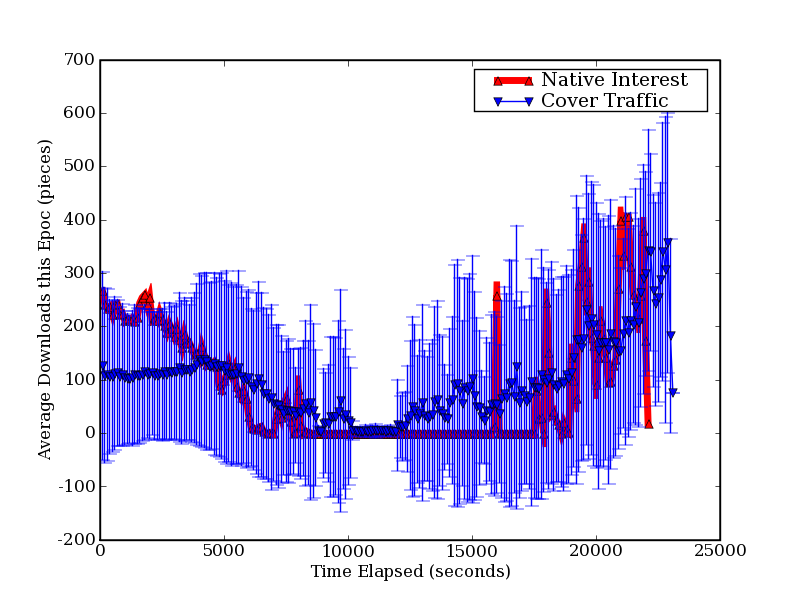}
\caption{Native and non-native traffic patterns super-imposed. While native traffic is above non-native traffic for the
same node, the median for the native is within the standard deviation of the 
other.\label{id1_composite_traffic}}
\end{figure}

\paragraph{Download Rates Over Time.} Finally, we examined
the rate of piece transmissions for native and non-native populations in the 
Marketplace to verify that transmission \textit{patterns}
are indistinguishable.
We created our transmission pattern by aggregating each node's download volume within
500 second buckets. All nodes are normalized such that their first 500 second slice
of time is slice 0, the second 500 seconds is slice 1, and so forth. Within each slice,
we separated the download volume for the native interest from the average download volume for
the cover-traffic. The average for all nodes and the standard deviations are computed
for each time slice. Figure~\ref{id1_composite_traffic}
shows the download pattern for all nodes across the entire simulation.
We again observe that the nodes' averages for native traffic is within the
standard deviation of the cover-traffic. Note also that this graph represents
a global view over all nodes, so this any node's local view would have higher error.

\begin{figure}
\centering
\includegraphics[scale=.30]{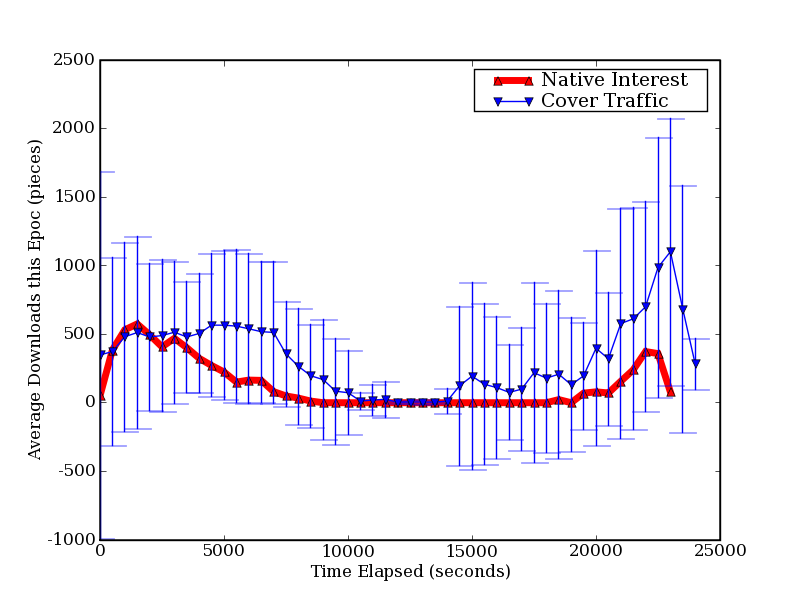}
\caption{This figure is similar to Figure~\ref{id1_composite_traffic} but limited to
the viewpoint of single clients. In other words, the former figure is a global representation
of download patterns, while this figure is representative of what a single peer observes.
\label{id1_clientview}}
\end{figure}

We can examine a weaker observer by computing the observed download patterns
for a single client. That is, for each node, we aggregated all the traffic that only
that node observed directly. As before, we aggregated into 500 second buckets, dividing
the native interest traffic from cover traffic. Then we used the average and standard deviations
for each node's observed patterns to create Figure~\ref{id1_clientview}. The two types of
traffic overlap even more in this graph, demonstrating that a single peer observes
less differences between native interest traffic and cover traffic then can be observed
across the swarm as a whole.

\subsection{Analysis}
We now revisit anonymity against each of the inquisitors that we previously identified.

\paragraph{Passive Inquisitor.} These nodes do not directly interact with any actual 
nodes but only talk to the tracker or DHT. The passive inquisitor can,
at best, track a given node's active set. From this information, it cannot determine the
node's native interest. As we demonstrated, the entrance and
exits of a given torrent in a node's active set appear indistinguishable, regardless of the torrent's
status as native interest.

\paragraph{Decoy Passive Inquisitor.} These nodes directly interact with other nodes,
but do not actually exchange pieces. They can, however, advertise pieces and unchoke
other nodes. Such inquisitors gain additional information, 
because rational nodes will drop them regularly for their poor performance.
However, with a Sybil attack~\cite{sybil}, these nodes can connect to a given node over and over
from different IP addresses, simulating a continuous connection. Such a Sybil attack
could track the traffic of a rational node by capturing all \emph{Have} announcements.
Nevertheless, even a Sybil attacker will not determine the node's native interest from this information
because, as we demonstrated, the download rates for a given
torrent for a node are similar, regardless of the node's native or non-native interest in that torrent.

\paragraph{Active Inquisitor.} The most powerful non-wiretap node, these nodes actively trade
with peers in the network. This feature allows them to attempt to ``trick'' a victim node
into revealing state through carefully crafted trading. For example, an active inquisitor might obtain a large
number of blocks from all the nodes in active set. Then, it might selectively advertise these
blocks to the victim to see which blocks the victim takes a higher interest in. Furthermore, a very
well provisioned inquisitor might introduce \textit{identifiable torrents} into the marketplace that
it can use to manipulate torrent values within a neighborhood. The active inquisitor can use
such value manipulation to attempt to pierce the indistinguishability.

At present, we have not yet attempted to simulate active inquisitors. Nevertheless, we expect
that unless the inquisitor can control a large portion of a victim node's local neighborhood (e.g.,
using a Sybil attack), it cannot have high confidence about the motivation for a node's
interest in any given torrent. This attack, however, is made non-trivial because DHTs or trackers give out
random subsets of the peers to a participating node, thus dramatically increasing the costs of overtaking a node's
neighborhood. Nevertheless, Sybil attacks are a significant
security issue and remains a point of research.

In addition to our successful anonymity results, we also quantified the costs in these simulations.
The amount of data downloaded, expressed as a multiple of a single torrent, averaged $5.71\pm0.43$
Given that the optimal value is $5$, this indicates that our nodes are not wasting a lot of time downloading
torrents that they do not complete.

To conclude our evaluation, we review our incentives qualitatively along two of three axes suggested by previous
work~\cite{faithfulness-formal, faithfulness-applied}.
We now consider incentives for communication and incentives for computation.
There is no need to evaluate incentives for message passing because the Marketplace, as in
regular BitTorrent, does not have peers relay messages for one another.

\paragraph{Incentives for Communication.} The first question is, \textit{does a rational node
have any incentive to lie about its state?} 
\begin{enumerate}
  \item \textbf{Active Torrents}: The only incentive for a node to lie about its active set is for
  increased anonymity against passive inquisitors. However, we have demonstrated that
  the native interest of a node is not revealed by the makeup and dynamics of the active set.
  Furthermore, the active set is necessary for performance and anonymity. Therefore, there
  is no incentive to lie about this state.
  \item \textbf{Choke Status}: There is no incentive for a rational node to misinform a neighbor
  about the choke state between them. A lie about choke status might result in a snub,
  which is undesirable.
  \item \textbf{Piece-Interest Status}: The incentives to lie about this are unclear.
  There is an incentive for a node to announce that it has pieces,
  even for pieces it does not actually have because the value of the torrent in the marketplace will increase.
  On the other hand, unchoked neighbors may ask for these pieces and subsequently snub
  the lying node when it cannot produce them. We have not yet quantified these incentives,
  but snubing is undesirable, providing a disincentive to this behavior.
  \item \textbf{Piece Requests}: A node has an
  incentive to request pieces that it already has in order to drive the value of the torrent higher.
  However, it also affects the value of torrents by pretending not to have the piece.
  Requesting pieces already present costs additional bandwidth, which is valuable and limited,
  so that behavior is certainly disincentived.  Similarly, pretending not to have a piece means
  that a peer who might have something to trade might skip over this node.  As with core BitTorrent,
  Marketplace nodes have an incentive to participate normally in torrent trading such that they
  get what they want in an efficient manner.
\end{enumerate}

\paragraph{Incentives for Computation.} We next ask, \textit{does a rational node
have any incentive to compute a non-conforming value for torrents in the marketplace?}
The answer is no, by definition, because nodes will compute their own market valuations. 
Theoretically, all nodes have an incentive to develop effective methods of evaluating torrents
of non-native interest. The cooperation model supports and encourages this form of 
self-interested operation.

In summary, the Marketplace is built on a sound foundation of incentives, although some 
small components are currently manipulable, and aggressive Sybil attacks may be
able to weaken the anonymity guarantees. These are open problems for future research.

\section{Discussion and Future Work}
\label{discussion}
Our proposal of the BitTorrent Anonymity Marketplace is a valuable
contribution to p2p-anonymity, particularly if an implementation of it
could draw away traffic from Tor. However, our work has produced many
more questions than it has answered.

\subsection{Stronger Anonymity and Ethical Issues}

Our anonymity model is designed to shroud a peer's intentions from
the observations of its neighbors. However, many BitTorrent users would
be interested in shrouding their intentions from adversaries that can
tap their wire, such as their ISP.
The BitTorrent Anonymity Marketplace could potentially  be hardened to improve
anonymity in such cases when the adversary can tap the peer's line.

\paragraph{Per-peer encryption.} Peers can communicate with one another
via encrypted links, an optional feature already present in BitTorrent. This immediately hides the message exchanges that
divulge the Marketplace's state. Despite this link encryption,
an adversary would still have access to the public information in the DHT.

\paragraph{Late-start native interest.} A node does not need to connect to
its native interest upon initialization. Instead, it can choose its $k$-active
set randomly, which may or may not include the native interest.
If not present in the initial active set, the node can rotate it into activity
at a later time.

\paragraph{Even split peers.} Because our system biases a node's selection
of its peers based on the value of the torrents they are trading, an observer
could approximate the value of each torrent to the node based on its neighbor
selection.  Nodes could remove this bias, selecting peers evenly from their
desired torrents.


\paragraph{Improving cover traffic.} Users that are sensitive to their
anonymity should ensure that the Marketplace is well stocked with items
that are legitimate candidates for cover traffic. Such items would include
sensitive, but highly-legal objects that provide better plausible deniability.


This last point about cover traffic leads to interesting 
ethical questions about the BitTorrent Anonymity Marketplace because it will, without a doubt, provide
cover for individuals engaging in illegal and reprehensible behavior. Unfortunately,
it is often the assumption that anonymity \textit{only} benefits individuals engaging
in such actions.
The truth is that anonymity is valuable for many legitimate purposes. For example,

\begin{itemize}
\item An individual with a medical condition may not wish to reveal it. Doing research
on the internet can expose them to other parties. The BitTorrent Anonymity Marketplace
does not provide anonymity for the initial search for documents (a standard service like
Tor is well suited to this task), but could provide cover for 
downloading and viewing a video about treatment options.

\item Legality is highly dependent on the jurisdiction. What may be legal in one region of the 
world may be highly illegal somewhere else. Such content may be sensitive
to the downloader even if it is legal. This is especially true if the downloader is from, or has
ties to, a jurisdiction where it is illegal.

\item Anonymity also protects individuals from commercial exploitation. In cases where
BitTorrent is being used for legal content, corporations can easily learn a user's
tastes and interests from very simple observations of the tracker or DHT. Absent regulations
to the contrary,
corporations will naturally begin using this information to target users with
advertising and so forth. The BitTorrent Anonymity Marketplace significantly reduces
the effectiveness of such attacks, since many or most of the nodes participating
in any given torrent will be there for the cover-traffic, not because it's their native interest.
In fact, they will have no idea what they're sharing.
\end{itemize}

The effectiveness of the Marketplace is greatly increases when there are many kinds of
legitimate, yet sensitive, torrents actively in trade. On the other hand, if only illegally
copied music is found therein, it won't matter if you have $k$-anonymous cover traffic. $K$ illegal music
or movie downloads is no better (and, in fact, could be worse) than just one.

That said, there will be individuals that would be interested in using a service such as the BitTorrent
Anonymity Marketplace to engage in illegal behavior. They should be aware that $k$-traffic
anonymity will probably not shield them effectively from government observation (see, e.g.,
You-are-not-a-lawyer~\cite{YANAL}). It is possible, 
however, that the BitTorrent Anonymity Marketplace \textit{does} help to cover users
against corporate investigation. For corporations looking to bring lawsuits against individuals
based on downloads, the BitTorrent Anonymity Marketplace greatly increases the cost
of determining infringement, and introduces a risk of false positive to the suing company.

Can a user be held legally liable for downloading a torrent, as cover traffic, assuming the
content in question would be illegal to have downloaded via ordinary means?  The essence
of the user's defense would be that they were just helping random peers to download content,
while they, themselves, were getting something entirely different.  Of course, if they are faced
with all $k$ of their encrypted downloads and asked to prove which one they can decrypt,
they may be stuck. Furthermore, even if the user legitimately doesn't know what is being
downloaded, the adversary might well crawl the various content discovery sites (e.g., PirateBay
and the like), creating their own reverse-mapping from encrypted torrents to their true identities.

As such, the degree of anonymity proffered by the BitTorrent Anonymity Marketplace seems
to be comparable to serving as the exit node of Tor or another such onion-route system.
The exit node is clearly observable doing fetching what could well be illegal content.  The
exit node's operator may well claim that the content in question was being delivered to a
third party, but the exit node is clearly participating in the process.  Of course, such arguments
quickly become absurd.  Internet core routers certainly have significant volumes of undesirable content transiting
them every day, all day long.  They might claim a ``common carrier'' defense if sued.  Could
a BitTorrent Anonymity Marketplace node, or for that matter a Tor exit node, claim a similar defense?

\subsection{Informed Risk}

One possible development to the BitTorrent Anonymity Marketplace would be channels that
inform the participant of risk. In particular, these third-parties would uncover the content
names and descriptions for the opaque DHT identifiers. Users of these services could then
fill their active sets with elements from white lists or prevent elements from black lists from
getting in. This would, of course,
erase plausible deniability about not knowing the content. However, the user could choose
their own level of risk.

Most importantly users could be absolutely sure that morally, ethically, and legally
unacceptably risky content, such as child pornography, would never pass through their
systems. Users looking for anonymity for sensitive but legal content, such as medical treatment videos,
could also ensure that they were not taking any legal risks for their behavior and might, instead,
find themselves downloading medical videos for a wide variety of different ailments. Moreover, certain
organizations that believe in civil disobedience to what they perceive as unjust laws might 
purposefully participate in providing cover traffic for certain classes of torrents. Curiously, the black list
for one organization might be a white list for another.

As a concrete example, consider a government that runs a black list of videos that are 
deemed illegal for whatever reason (e.g., criticism of the king is illegal). Citizens within that country that wish to have anonymity
and avoid legal risk could use that list as a black list. Other individuals, inside or outside of the country,
might treat that as a white list, looking to provide cover traffic for those torrents by making them
more popular.

\subsection{Future Work}

Several aspects of the BitTorrent Anonymity Marketplace remain unresolved or require further exploration.
The aforementioned legal issues are one such area. It would be valuable to explore the legal 
possibilities of the BitTorrent Anonymity Marketplace under the laws of various jurisdictions.

Another area of significant future research is the valuation function that each peer performs on the
torrents it is trading. Just as we are not lawyers, we are also not economists. We recognize that the
economic interactions of our proposed system are complicated but subtle. In a real world implementation,
there might be thousands of torrents and hundreds of thousands of clients in the Marketplace, not to
mention churn, disparities of upload and download capacities and so forth. It will be a daunting challenge
to uncover a generalized valuation function that works well under all circumstances.

Our current simulations are pedagogical and unrealistic. In particular, we have not studied the BitTorrent
Anonymity Marketplace under realistic churn or other such conditions. Because our simulations lack these
features, we have been unable to see some predicted behaviors that require them. Also, in a real-world scenario,
torrents will be of different sizes and nodes would have widely varying network performance.  Different nodes
might have different values of $k$-anonymity that they desire.
It would be convenient if the choice of $k$ value for a client had no impact on its neighbors, but we have not examined this.

We have also not completely explored the attack space for either inquisitors or rational attackers. Our simulation
does not yet include an active inquisitor that attempts to introduce tainted information in an effort to 
reveal the interests of peers. Similarly, our simulations do not yet include a rational manipulators that lies about
state in an effort to manipulate torrent values. 

Finally, it should be obvious that simulation alone is insufficient for evaluating the BitTorrent Anonymity
Marketplace. An actual implementation must be created and evaluated for real-world operations. A whole
host of difficulties is involved in such development, although most of them are legal, rather than technical.

\section{Conclusion}
\label{conclusion}

In this work, we have explored a new method for cooperative 
anonymity in BitTorrent swarms, called the
BitTorrent Anonymity Marketplace, where peers exchange pieces of
multiple torrents based on their value for trading with other peers. This
creates a world where intent is difficult to discern because motivations
are obscured by the shifting values within the local neighborhood.
Nodes always download $k$ different torrents, selected randomly,
to completion, obscuring their true intent, yet still biased in favor
of increasing the nodes' observed performance.

With detailed event-based simulations, we demonstrated that  
the download behavior for native interests and cover traffic
was statistically similar, making it difficult for observers to distinguish between
the two. We also demonstrated in simulation that our Marketplace completes
without unreasonable overhead beyond the cover traffic's costs. We also evaluated the incentives of
our system and found that the overall setup is sound against rational manipulations, but that there are 
obvious places for exploitation.


\section*{Acknowledgements}
Thanks to Gabriel Landau for suggesting the phrase ``indistinguishability of intent.''

Additional thanks to our colleagues at Independent Security Evaluators for feedback, reviews, and suggestions.

{\small
\bibliographystyle{abbrv} \bibliography{bam,peer2peer,proposal,twngan,attackstaxonomy,prior_work}
}

\end{document}